\documentclass[twocolumn]{aastex631}  
\usepackage{relsize} 
\usepackage{graphicx}	
\usepackage{amsmath}	
\usepackage[flushleft]{threeparttable}
\usepackage{changepage}
\usepackage{epstopdf}
\usepackage{amsopn,amsxtra,txfonts}
\usepackage{comment} 
\usepackage{longtable,threeparttablex}
\usepackage{enumitem}
\usepackage{fancyhdr}
\usepackage{booktabs}
\usepackage[hang,flushmargin]{footmisc}
\usepackage{gensymb}
\usepackage{tabularx}
\usepackage{ragged2e}
\usepackage{float}
\usepackage{enumitem}
\usepackage{hyperref}
\usepackage{wrapfig}
\usepackage{sidecap}
\usepackage{tcolorbox} 
\usepackage{xcolor} 
\usepackage{natbib}
\usepackage{threeparttable}
\usepackage{array}
\usepackage{booktabs}

\newcommand{\OI}{\ion{O}{1}}
\newcommand{\NI}{\mbox{N\,{\sc i}}}

\newcommand{\FeII}{\mbox{Fe\,{\sc ii}}}

\newcommand{\SiII}{\mbox{Si\,{\sc ii}}}
\newcommand{\SiIII}{\mbox{Si\,{\sc iii}}}
\newcommand{\SiIV}{\mbox{Si\,{\sc iv}}}
\newcommand{\CII}{\mbox{C\,{\sc ii}}}

\newcommand{\CIV}{\mbox{C\,{\sc iv}}}

\newcommand{\siii}{\ion{Si}{2}}
\newcommand{\siiii}{\ion{Si}{3}}

\newcommand{\ovi}{\ion{O}{6}}
\newcommand{\ovii}{\ion{O}{7}}
\newcommand{\neviii}{\ion{Ne}{8}}
\newcommand{\hi}{\ion{H}{1}}

\newcommand{\nhi}{$N_{\rm H\,I}$}
\newcommand{\lya}{Ly$\alpha$}

\newcommand{\kms}{\,km\,s$^{-1}$}

\defcitealias{Danforth16}{D16}

\begin{document}

\title{Discovery of Weak \ovi~ Absorption in Underdense Regions of the Low-Redshift Intergalactic Medium}
\author[0000-0002-4157-5164]{Sapna Mishra}
\affiliation{Space Telescope Science Institute, 3700 San Martin Drive, Baltimore, MD 21218, USA}

\author[0000-0002-8433-5099]{Vikram Khaire}
\affiliation{Department of Physics, Indian Institute of Technology Tirupati, Tirupati, Andhra Pradesh 517619, India}

\author[0000-0000-0000-0000]{Romeo Pallikkara}
\affiliation{Department of Physical Sciences, Indian Institute of Science Education and Research (IISER), Mohali, Punjab 140306, India}


\author[0000-0000-0000-0000]{Anand Narayanan}
\affiliation{Indian Institute of Space Science \& Technology, Thiruvananthapuram, Kerala  695547, India}


\author[0000-0003-0724-4115]{Andrew J. Fox}
\affiliation{AURA for ESA, Space Telescope Science Institute, 3700 San Martin Drive, Baltimore, MD 21218}

\correspondingauthor{Sapna Mishra}
\email{smishra@stsci.edu}

\begin{abstract}

We search for weak \ovi\ absorption in the low-redshift intergalactic medium (IGM) using 82 high signal-to-noise quasar spectra from the Cosmic Origins Spectrograph on the Hubble Space Telescope. From this dataset we compile a clean sample of 396 intervening Lyman-$\alpha$ (\lya) lines with \hi\ column densities log\,($N_{\rm H\,I}$/cm$^{-2})<14.5$ and no individual \ovi\ detections. Stacking at the location of the \ovi\ doublet reveals absorption at $>5\sigma$ significance, with equivalent width (W$_r^{1032}) =1.7\pm0.3$ m\AA, corresponding to log\,($N_{\rm O\,VI}$/cm$^{-2}$) = $12.14\pm0.08$. The stacked \ovi\ signal associated with strong \lya\ (13.5 $\leq$ log\,$N_{\rm H\,I}<$ 14.5) absorbers is significantly stronger than that associated with weaker \lya\ (12.5 $\leq$ log\,$N_{\rm H\,I} <$ 13.5) absorbers. For the subset of 81 broad \lya\ absorbers (BLAs; $b_{\rm H\,I} > 45$ km s$^{-1}$), we obtain a marginal $\sim4\sigma$ \ovi\ detection. Other than \siiii, detected at 5$\sigma$, no metal lines are found. Cross-correlation with galaxies shows that 93\% of \lya\ absorbers lack an associated bright galaxy ($>3L^\ast$) within 1~Mpc. A more complete low-redshift ($z<0.25$) search shows 66\% \lya\ lack galaxies down to $L^\ast$. This suggests the \ovi\ arises mainly from the diffuse IGM, not the circumgalactic medium. The stacked \ovi\ signal suggests characteristic metallicities of $\approx 0.01Z_\odot$ under photoionisation and $\approx 0.001Z_\odot$ under collisional ionisation, though these estimates are model-dependent and assume \ovi\ and \hi\ trace the same phase. This study provides the first observational evidence for metal absorption in low-column-density \lya\ systems that individually show no detectable metals, placing important constraints on metal enrichment of the underdense IGM.

\end{abstract}


\section{Introduction}

A comprehensive census of the baryon inventory in the low-redshift universe reveals that 
observations fall short of expectations, and only account for 30--40\% of the baryons 
predicted by big-bang nucleosynthesis 
\citep{Fukugita1998, Lehner07, Bregman2007, Shull12_baryons, Planck20}.
Although fast radio burst observations have recently closed the total baryon budget \citep{Macquart2020, Connor2025}, 
the exact phase in the temperature and density plane of these baryons remains unclear.
Simulations of cosmological structure formation \citep[e.g.,][]{Dave01, Martizzi19_tng_whim, Tuominen21} suggest that a significant portion of the missing baryons exists in the intergalactic medium (IGM) consisting a diffuse, low-density ($n_{\rm H} \sim 10^{-6}$ -- 10$^{-4}$ cm$^{-3}$) warm-hot phase \citep[$T \sim 10^{5-7}$ K;][]{Cen1999}, referred to as the warm-hot intergalactic medium (WHIM).
A large fraction of this WHIM gas cannot be probed with Lyman-$\alpha$ (hereafter, \lya) absorption lines since the neutral fraction of \hi~is low ($f_{\rm H\,I} \sim$ 10$^{-7}$ to 10$^{-5}$) and / or the lines are too broad (with Doppler parameter $b_{\rm H\,I} >$ 45 \kms) to imprint significant absorption on the 
spectra \citep[e.g.][]{Narayanan2010,Tepper-Garcia12, Hu23_whim}.
However, there are individual detections of such broad \lya~absorption (BLA) lines, 
identified via asymmetry in the absorption line profile of nearby strong 
\lya~absorbers \citep[see e.g][]{Richter06, Savage11, Narayanan12, Pachat16}.
These individual detections probe only $\sim 10$\% of the total 
baryons predicted in the WHIM \citep[][]{Danforth10, Shull12_baryons, Tejos16}. 

However, there is an alternative approach for investigating the WHIM, 
involving the detection of
highly ionized metal species such as 
\ovi~\citep[e.g,][]{Tripp2000, Tripp08, Savage14, Werk16_ovi}, 
\ovii~\citep[][]{Nicastro19}, or \neviii~\citep[][]{Savage05, Narayanan2009, Meiring13, Pachat2017, Burchett18}. 
These ions serve as effective probes of gas with temperatures above $10^{5}$ K when 
the dominant ionization mechanism is collisional ionization. 
A significant challenge in these metal absorbers 
lies in determining the ionization mechanism, as studies have indicated that some of 
these metals may also arise from low-density photoionized gas 
\citep[e.g.][]{Hussain16, Hussain17, Chen2017}. Moreover, accurate modeling of metal ions also crucially depends on their assignment to single or multiple phases \citep{Sameer2021, Sameer2024b, Haislmaier2021}. 
Therefore, the interpretation of these observations hinges on the 
uncertain spectral slopes of extragalactic UV background \citep[see][]{Khaire2015,KS19, Acharya2022} 
at the ionization potential of these ions. 

Searching for highly ionized gas is important not only for probing the WHIM but also for assessing the metal content within the IGM 
and the circumgalactic medium (CGM). Previous surveys targeting low-redshift
\ovi\ absorption have provided valuable 
insights \citep[e.g][]{Tripp08, Thom2008a, Thom2008b, Danforth2008, Savage14, Werk16_ovi}, yielding metallicity measurements of the \ovi-bearing gas 
ranging from $\approx0.01-1\,Z_{\odot}$ \citep{Savage14, Sameer2024}.
These surveys primarily focused on detections of individual \ovi\ absorbers, in most instances with strong associated \lya~absorption with high \hi~column densities
(log\,$N_{\rm H\,I}>$14.5). \ovi~is known to be prevalent in high \hi~column density clouds due to their vicinity to galaxies and higher oxygen abundances  
\citep[e.g.][]{fox2013, Werk16_ovi, Sameer2024},  
potentially biasing the metallicity estimates for these detections toward higher values, such that they may not reflect the metallicity distribution of the broader \ovi-bearing IGM.

Hence, it is important to extend the search for \ovi~to low-\hi\ column density systems
(log\,$N_{\rm H\,I}<$14.5), which trace the more diffuse regions of the IGM where the WHIM is predicted to reside. 
However, detecting individual \ovi~absorbers in such systems is challenging, as the absorption features will be significantly weak, concealed within the noise of the spectra. 
The limited signal-to-noise (S/N) of individual sightlines prevents a direct detection of these weak absorbers. 
To overcome this limitation, we make use of the spectral stacking technique \citep[see e.g.,][]{Frank2018,Lan2018,Mishra2022, Yang2022, Mishra2024}, which combines the spectral regions corresponding to the expected \ovi~ doublet from a large ensemble of \lya\  systems. 
By co-adding these spectra in the rest frame of the absorbers, random noise averages out; at the same time, any \ovi~absorption signal is reinforced, thereby enabling the detection of statistically significant but individually undetectable absorption features. 
This approach provides a means for measuring the average metal content of the low-density IGM and extending the census of intergalactic metals to regimes inaccessible through individual line detections.

In this study, we explore the high-quality dataset of \citet[][hereafter 
\citetalias{Danforth16}]{Danforth16}, focusing on the low-redshift IGM. This dataset 
contains 82 quasar spectra observed with the FUV gratings of the Cosmic Origins Spectrograph 
(COS) on board the Hubble Space Telescope (HST), with a high S/N of more than 
10 per resolution element. We utilize the absorption line identification (IDs) provided
by \citetalias{Danforth16} and select a clean, uncontaminated sample of 
396 \lya~absorption lines where there is no individual 
detection of corresponding \ovi~absorption with log\,$N_{\rm O\,VI} >13$. On this sample, we perform the spectral 
stacking using three different methods: median, S/N weighted mean, and 5$\sigma$ clipped mean. 
All three methods yield a clear detection of weak \ovi~doublet at more than 5$\sigma$ statistical significance. 
This represents the first detection of its kind. This paper is organized as follows: In Section~\ref{sec.data}, we outline our approach to 
sample selection. Section~\ref{sec:stack_analysis} describes the stacking analysis methods and Section~\ref{sec:results} presents the results obtained from these methods. In Section~\ref{sec:discussion}, we discuss the galactic association and metallicity contributed from the detected \ovi~absorption and its implications on 
the outcomes. Finally, in Section~\ref{sec.results}, we provide a concise summary of our findings. Throughout the paper we use a flat $\Lambda$CDM cosmology with $H_{0} =$ 70 km s$^{-1}$ Mpc$^{-1}$ , $\Omega_{M} =$ 0.3, and $\Omega_{\Lambda} =$ 0.7.

\section{Sample Selection for \ovi\ Search}\label{sec.data}

We analyze the \citetalias{Danforth16} low-z IGM dataset containing 82 high S/N ($>10$ per resolution element) HST COS quasar spectra observed with G130M and G160M gratings. Most of these quasars are observed under Guaranteed Time Observation programs (PI Green), and some are under Guest Observer programs. The spectra were reduced and co-added by \citetalias{Danforth16}. The final spectra with all the 
line IDs are provided by \citetalias{Danforth16} on the MAST webpage\footnote{\href{https://archive.stsci.edu/prepds/igm}{https://archive.stsci.edu/prepds/igm}}. 

The \citetalias{Danforth16} quasars probe the \lya~forest at $z < 0.48$. 
However, our search for \ovi~and corresponding \lya~lines confines us to a limited redshift 
range, $0.09<z<0.48$. This redshift range ensures the spectral coverage of both 
\ovi~doublets and \lya~lines within the COS spectra, spanning wavelengths from 1122 \AA~to 
1800 \AA\ as covered by G130M and G160M gratings. Within this redshift interval, \citetalias{Danforth16} identified a total of 1875
\lya~lines. Among these, 436 exhibited individual \ovi~detections (with at least one line of the doublet) 
with log\,$N_{\rm O\,VI}>13$ as reported in the same catalog 
with a tolerance of $\Delta v = 50 $ km s$^{-1}$ around the expected \ovi~locations. This tolerance window is chosen to account for the typical velocity offset observed between low-redshift \lya~ and \ovi~ absorbers in the literature \citep{Tripp08, Thom2008b, Savage14}. We excluded these \lya~lines, resulting in a sample of 1439 \lya~absorption lines without individual \ovi~detection. 
Subsequently, for completeness and to avoid inclusion of any misidentified \lya~line, we also removed 132 \lya~lines lying in the geo-coronal \lya~line region ($\sim$ 1208--1224 \AA). This resulted in 1307 \lya~absorption lines.

For these 1307 \lya~absorption lines, we searched for contamination from any other lines, including intervening absorption and strong
interstellar medium absorption originating from our Milky Way 
(such as \SiII, \SiIII, \FeII, \CII, \SiIV, \CIV) at the expected locations of \ovi\, $\lambda\lambda1032,1038$ doublet lines.
For achieving this, we followed the line finding algorithm outlined in \citetalias{Danforth16}. To determine individual absorption lines, \citetalias{Danforth16} introduced a significance 
level (SL) based on equivalent width, defined as 
$SL(\lambda) = W(\lambda)/ \bar{\sigma}(\lambda)$, where $W$ represents the equivalent width of 
the line and $ \bar{\sigma}(\lambda)$ 
accounts for the error vector convolved with a Gaussian having a Doppler parameter of 
$b=20$\kms\ to approximate the COS Line Spread Function (LSF). 
Spectral regions with $SL \geq 3$ were identified as absorption lines. 
For each of these systems, we isolated a rest-frame 
wavelength region spanning from 1030--1040 \AA, 
encompassing the \ovi~ doublet, 
and conducted a careful search for potential contamination at the expected locations. To find contamination, we utilized the 
reported $b$ values of the \lya~lines in the \citetalias{Danforth16} catalog and 
examined if any other absorption lines that were detected within a range of $3$ times the Full Width at Half Maximum (FWHM) of the \lya~line (FWHM = 2$\sqrt{\ln 2} \; b$). 
We excluded 721 lines that 
exhibited contamination from other absorption lines. Excluding these 721 \lya~ lines yielded a clean sample of 586 \lya~ absorption systems, free of any detected \ovi~ absorption or contamination from other lines at the \ovi~ wavelengths. 

Finally, to detect \ovi~absorption from the diffuse IGM, we considered only systems with \lya\ column densities of ${\rm log}\,N_{\rm H\,I} < 14.5$. The majority of our \lya\ lines at this stage satisfy this criterion, except for seven systems. We therefore excluded these seven from our analysis, resulting in a sample of 579 clean weak \lya\ absorption systems.
We also required the spectral regions used for detecting the \ovi\,$\lambda\lambda1032,1038$ doublet to be free of geocoronal contamination. Therefore, we excluded all \lya\ systems for which the expected \ovi\ doublet 
absorption falls within the geocoronal emission regions of \NI\ ($1198$--$1202$\,\AA), \lya\ ($1208$--$1224$\,\AA), or \OI\ ($1300$--$1308$\,\AA). In addition, we retained only those systems that have at least five pixels within $\pm 50\,\mathrm{km\,s^{-1}}$ of both \ovi\ doublet lines. This criterion removed an additional 163 \lya\ systems, leaving 416 \lya\ absorbers. A subsequent visual inspection of these 416 absorbers at the expected \ovi\ doublet locations led to the removal of 20 more systems that showed absorption at the weaker \ovi~$ 1038$ line but no corresponding feature at the stronger \ovi~$ 1032$ line. After implementing these exclusions, the final, clean sample comprises 396 \lya\ absorbers.

\begin{figure}[!ht]
      \includegraphics[width=0.42\textwidth]{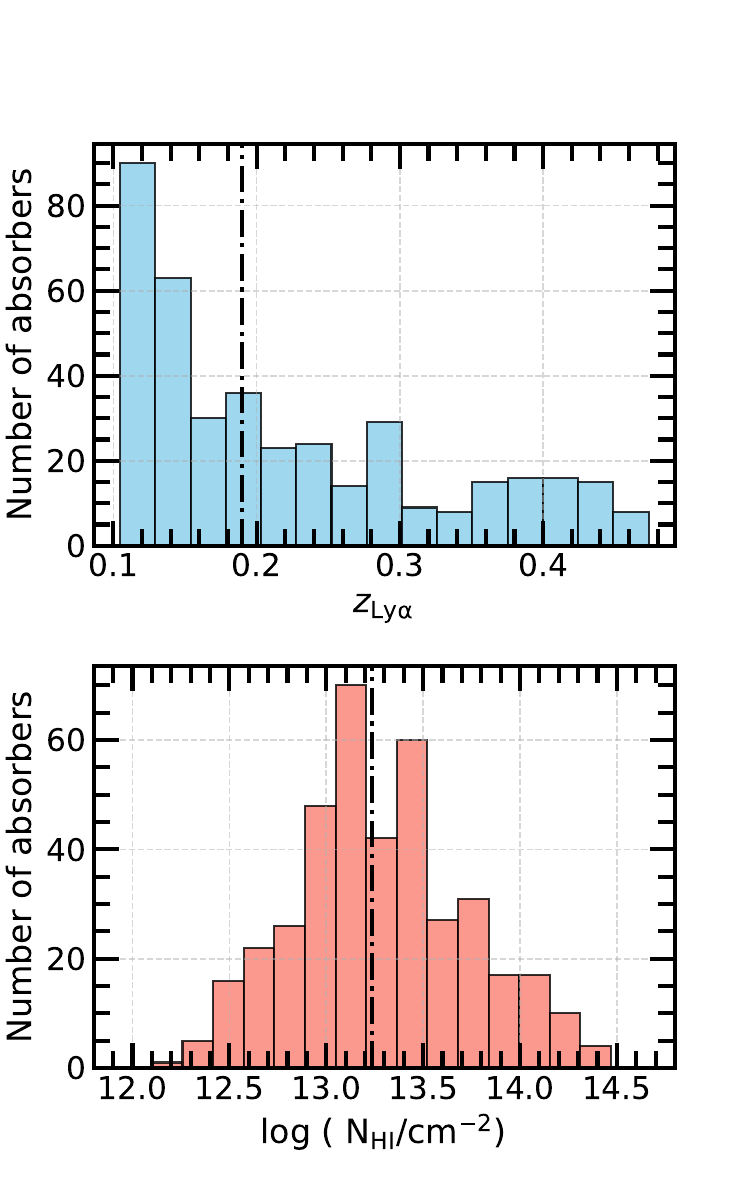}
     \caption{Histograms showing the redshift (top) and column density (bottom) distributions of our final, clean 396 weak \hi\ absorbers that defines our stacking sample to detect weak \ovi~ IGM absorption. The dotted-dashed line in each panel marks the median of the distribution.}     \label{fig:hist_HI}
    \end{figure}

In Fig.~\ref{fig:hist_HI}, we show the histograms of the redshifts (top) and column densities (bottom) of these 396 \hi~ absorbers tracing a subset of low column density \lya~forest region. This carefully selected sample served as the basis for our spectral stacking analysis, aimed at uncovering the presence weak \ovi. The specifics of our stacking methods 
are detailed in the following section.

\section{Spectral stacking analysis}
\label{sec:stack_analysis}

To perform the spectral stack, we first convert the observed wavelength into the rest-frame 
wavelength by dividing the wavelength array of each \lya~system by ($1+z_{\alpha}$) 
where $z_{\alpha}$ is the redshift of the absorber. Subsequently, we segment the wavelength 
intervals from 1030 to 1040 \AA~designated for stacking the \ovi~doublet. 
These segments are then rebinned with a consistent pixel separation of 0.03\AA, as 
initially defined in \citetalias{Danforth16}, 
corresponding to $\Delta v = 6$ km s$^{-1}$. 

In our stacking approach, we do not use the global continuum fits provided by 
\citetalias{Danforth16}. Instead, we opt for a more localized continuum fitting strategy, 
focusing on the 10\,\AA~segment of each spectrum to attain more accurate representations of 
the local continuum variations. For this purpose, we employ an automated code developed by \citet{Mishra2022}.
Briefly, we adopted the continuum from \citetalias{Danforth16} as an initial baseline continuum and normalized the spectrum. We then applied an iterative boxed sigma–clipping procedure with asymmetric sigma levels, chosen based on the median S/N of each spectrum, to remove absorption features while preserving residual emission. The number of boxes was adjusted depending on the presence of strong quasar emission lines. The iterative clipping was performed until most absorption features were removed. The clipped regions were then linearly interpolated, and a spline was fitted to the residual spectrum, yielding the final continuum.

After fitting the local continuum for all the \ovi~segments, we proceed to perform spectral 
stacking using three distinct methods as outlined in detail in \citet{Mishra2024}. To summarize, firstly, we implement the median stacking approach, 
wherein we compute the sample median at each pixel within our rebinned spectra.
Our second method employs S/N-weighted mean stacking. Initially, we calculate the S/N for each \ovi~segment by dividing the normalized flux by the normalized error 
vector and subsequently taking the median. In this S/N calculation, we exclude the region 
within $3 \times$ FWHM of the corresponding \lya~absorption line, as we anticipate the presence 
of \ovi~in this region. 
Following the S/N determination, we assign 1/(S/N) weights to each segment and 
calculate the weighted mean at each pixel. Our third stacking method involves a 5$\sigma$ 
clipped mean. Here, we compute a sample mean at each pixel while clipping any outliers that deviate from the mean by more than 5$\sigma$.
In this context, $\sigma$ denotes 
the standard deviation calculated from normalized flux values at each pixel. The results from these methods are shown in the next section.

\begin{figure}[!ht]
      \includegraphics[width=0.45\textwidth]{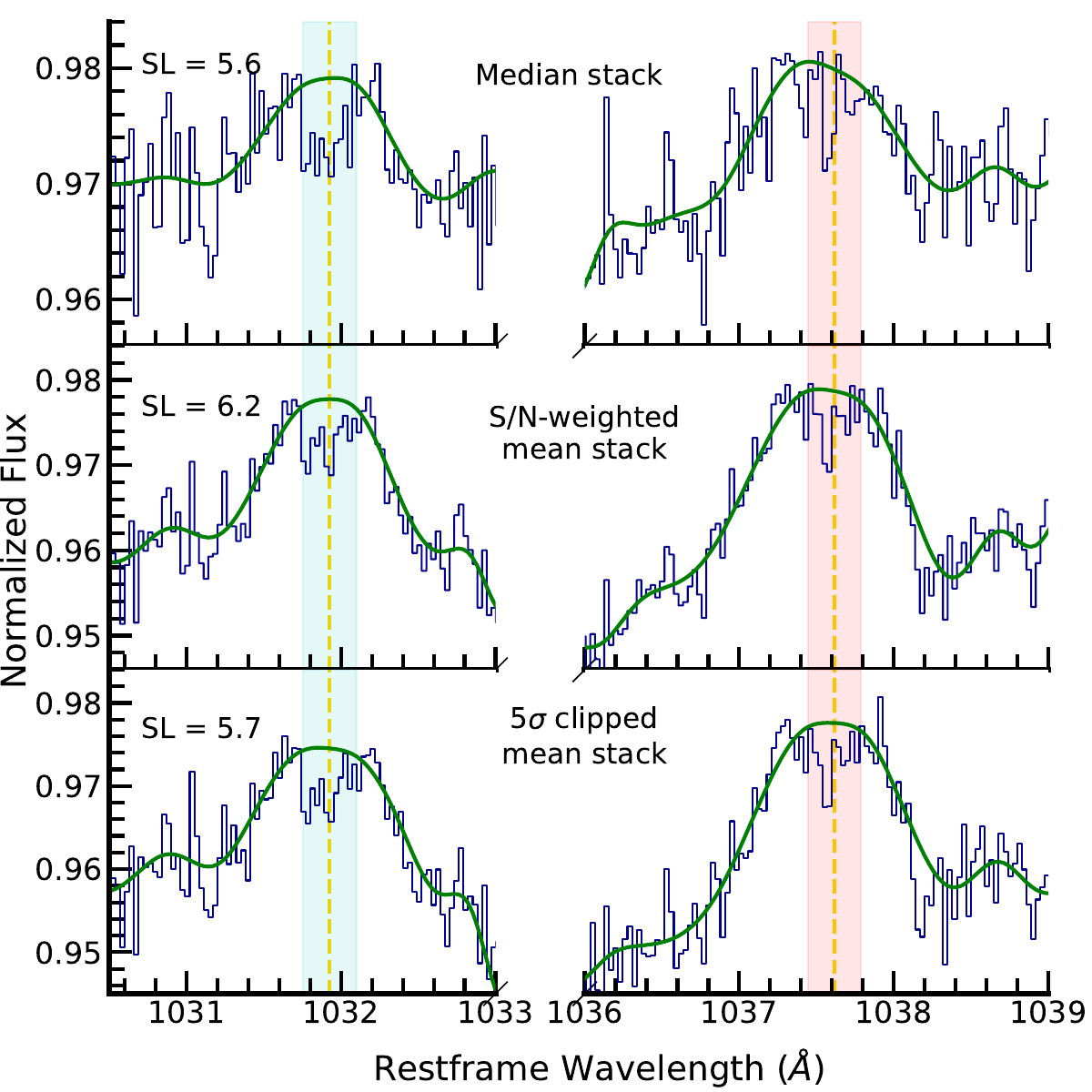}
     \caption{Stacked \ovi~absorption profiles obtained through three methods: median (top panel), S/N-weighted mean (middle panel), and 5$\sigma$ clipped mean (bottom panel). Vertical dashed lines represent the expected \ovi~doublet locations, while the shaded regions encompass the $\pm$50\kms~ around absorption features. The green curve in each panel shows the pseudo-continuum.  The emission-like bump in the left panels arises from the sample selection (see Section~\ref{sec:results}); removing this selection-induced feature in regions away from the detected \ovi~ absorption eliminates the bump (see Fig.~\ref{fig.appendix_sample_stack_no_bump}). All three methods yield \ovi~absorption with a statistical significance of $>5\sigma$ (legends indicate the combined significance of detection for both lines in each method, see Section~\ref{sec:results}).}
    \label{fig.sample_stack}
\end{figure}

\begin{figure}[!ht]
      \includegraphics[width=0.45\textwidth]{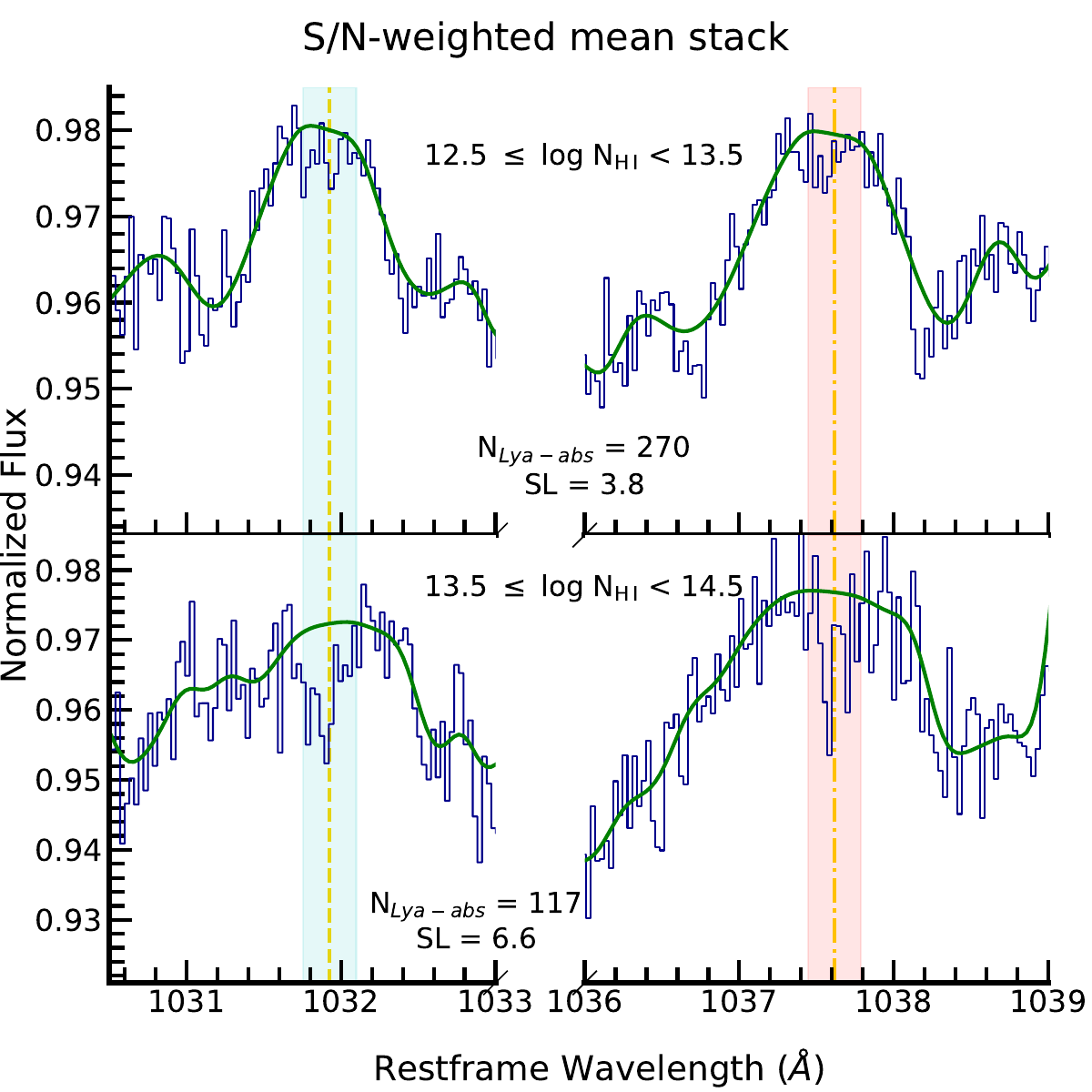}
     \caption{Similar to Fig.~\ref{fig.sample_stack}, this figure displays stacked \ovi~absorption profiles obtained using the S/N-weighted mean method, with results presented separately for two bins of \hi~column density: $12.5 < {\rm log}\,N_{\rm H\,I} <13.5$ (top panel, 3.3$\sigma$ detection) and $13.5 < {\rm log}\,N_{\rm H\,I} <14.5$ (bottom panel, 6.6$\sigma$ detection).}
     \label{fig.binned_stack}
    \end{figure}

\section{Results} 
\label{sec:results}

The stacked profiles from the three different stacking methods are shown in Fig.~\ref{fig.sample_stack}. The vertical dashed lines in each panel denote the expected positions of the \ovi~doublet. In all three cases, the spectral stacks exhibit clear absorption coinciding with the anticipated location of both lines of the \ovi~doublet. 
The expected \ovi\ absorption is superimposed on a broad, emission-like feature as can be seen in Fig.~\ref{fig.sample_stack} left panel. This broad emission-like feature arises because we ensured that the $\pm$50\kms\ shaded region around the \ovi\ doublet is free from any absorption. However, spectral regions beyond this window may contain unrelated random absorption, which can cause the stacked profiles to appear slightly suppressed in these regions 
\citep[as also shown in][]{Mishra2024}. We illustrate in Appendix Section~\ref{appendix:section_remove_uncor}, the emission-like feature disappears when uncorrelated absorption on both sides of the detected \ovi~ absorption feature is excluded (see Appendix Fig.~\ref{fig.appendix_sample_stack_no_bump}).

To mitigate these effects, we fitted a pseudo-continuum to these stacks. We employed the similar continuum fitting approach of iterative sigma-clipping and spline-fitting procedure used for the individual spectrum. The fit excluded the $\pm$50\kms\ region corresponding to the \ovi~doublet. The distinctive bump observed at the \ovi~doublet location warranted the use of more spline knots ($\sim$ 10) during the continuum fitting process. 
The resulting pseudo-continuum is shown by solid green lines in the left-hand panel of 
Fig.~\ref{fig.sample_stack}. \par

To calculate the significance of detection in the final stack, 
we rebin the stack to have the pixel size of the COS resolution element 
(i.e with $\delta v = 18$ km s$^{-1}$) by rebinning it by 3 pixels using the {\tt Python} routine {\it SpectRes} \citep{Carnall2017}. Then for the significance of detection of each line in the final stacks, we determined the significance level $SL$ of the absorption line by calculating 
$SL ^2  = \sum_{i} \big[(C_{p,i}-f_i)/\sigma \big]^2$ where $C_{p,i}$ and $f_i$ are the pseudo-continuum and flux at $i^{th}$ pixel
within $\pm 50$ km/s of the absorption line and $\sigma$ is standard deviation of flux excluding the region where we expect the \ovi.  
For the spectral stack including both lines of the doublet, we determine a combined significance of
$SL = \sqrt{ S_{1032}^{2} + S_{1038}^{2} }$,  which is 5.6$\sigma$ (5.0$\sigma$ for the blue and 2.5$\sigma$ for the red component) for our 
median stack, 6.2$\sigma$ (5.4$\sigma$ for the blue and 3.1$\sigma$ for the red component) 
for our S/N-weighted mean stack, and 5.7$\sigma$ (4.7$\sigma$ for the blue component and 
3.2$\sigma$ for the red component) for our 5 $\sigma$ clipped mean stack. 
In all three cases, the blue line is detected at $\approx 5\sigma$, while the red line is detected at $2.5$--$3.2\sigma$.
Consequently, the combined detection significance of the doublet across the stacks is $\approx 6\sigma$.
The measured rest-frame equivalent widths for the \ovi~1032 \AA\ (W$_{r}^{1032}$) are determined to be $2.0 \pm 0.3$~m\AA\ (median stack), $1.7 \pm 0.3$~m\AA\ (S/N–weighted mean stack), and $1.8 \pm 0.4$~m\AA\ ($5\sigma$–clipped mean stack). 
The doublet ratios (DR) measured for the detected \ovi\ absorption are $1.9 \pm 1.2$, $1.4 \pm 0.8$, and $1.3 \pm 0.7$ 
for the median, S/N–weighted mean, and $5\sigma$–clipped mean stacks, respectively. Within the quoted uncertainties, these values are consistent with the expected optically thin ratio of DR = 2.  
However, the relative strengths of the two components are sensitive to the placement of the local continuum due to the weak nature of the signal.
Besides the global pseudo continuum shown here, we tested alternative continuum fits (spline + polynomial, splines with varying knots across the spectral region).
We confirm different methods yield DR and EW values consistent within their quoted uncertainties.\par

Although our stacking methods provide means to recover weak absorption features, the approach carries potential caveats. 
The stacking procedure assumes that the “hidden” \ovi~ features are perfectly aligned with the corresponding \hi~ Ly$\alpha$ redshifts. Any systematic offset between the \ovi~and \hi~centroids (on the order of tens of \kms) will broaden and weaken the stacked signal, potentially leading to an underestimation of the stacked \ovi\ equivalent width. For higher–column-density \lya\ absorbers, the \ovi~and the strongest \hi~components are misaligned in as many as 60\% of cases \citep[see][]{Tripp08}. Indeed, such misalignments are expected in a multiphase medium. The “non-aligned’’ \ovi~typically occurs in systems where the \ovi~is tracing warm-hot gas with sufficiently high metallicity that the associated broad \hi~absorption becomes too weak to detect \citep{Savage2010}. There are also instances of low column density \lya~absorbers (log\,$N_{\rm H\,I} \geq 13.1$) where the associated \ovi~is offset in velocity from the \lya~ \citep{Williger2006}. Therefore, this possibility cannot be completely ruled out even for weak \ovi\ absorbers.\par

As also demonstrated in \citet[][see their Appendix B]{Mishra2024},
median stacking method may yield biased results when the S/N distribution 
within the data is not uniform. Conversely, $\sigma$-clipped stacking can lead to the removal of a substantial number of data points, introducing more noise into the stack. Given these considerations and the fact that our sample exhibits a skewed S/N distribution around the median S/N of $\sim 15$,  we opt to employ the S/N-weighted mean stacking method for our subsequent analysis and present the resulting measurements in Table \ref{tab:results}.\par

In addition to EW, we also provide the column density of stacked \ovi~using the linear part of the curve-of-growth (COG) using the 
standard equation 
\citep[e.g.][]{Petitjean1998}:
\begin{equation} 
   N^{\rm 1032} = 1.13\, \times 10^{20}\, \frac{{W^{1032}_{r}}}{f_{\rm 1032} \lambda_{1032}^{2}}\,\,\rm cm^{-2}
  \label{eqn:mg_col}
\end{equation}

Where $W^{\rm 1032}_{r}$, $f_{\rm 1032}$ and $\lambda_{\rm 1032}$ are the rest frame EW, oscillator strength, rest
wavelength of the \ovi~1032 line. Given the small values of EW (see Table~\ref{tab:results}, Column 6), the use of a linear COG is justified.
Additionally, in the Table~\ref{tab:results} we provide the median $z$, mean \nhi\ and mean $b_{\rm H\,I}$ values
from the sample using the reported values in \citetalias{Danforth16} catalogs. 

\subsection{Spectral Stacking for \lya\ Subsamples }
\label{subsec:bins}

To further investigate the origin of our detected \ovi~absorption, we divided our sample into 
two distinct bins based on \nhi~values: a high-column density sample having 
$13.5 < {\rm log}\,N_{\rm H\,I} < 14.5$, comprising 117 systems, 
and a low-column density sample 
having $12.5 <{\rm log}\,N_{\rm H\,I} < 13.5$, comprising 270 systems. We removed 9 systems 
with log\nhi~$<12.5$  when creating these bins. For both of these bins, we 
performed stacking on the \ovi~segments using our S/N-weighted mean stacking method and 
subsequently conducted pseudo-continuum fits, following the procedure outlined earlier. The 
resultant stacks are presented in 
Fig.~\ref{fig.binned_stack}. In the high-column density bin, we find a more pronounced 
\ovi~doublet as compared to the low-column density bin, featuring W$_{r}^{1032}$ of $3.6 \pm 0.7$~m\AA\ at 
statistical significance of 
6.6$\sigma$ (5.7$\sigma$ for the blue and 3.4$\sigma$ for the red) with DR of $1.3 \pm 0.7$.
The low-column density bin exhibits
W$_{r}^{1032}$ of $1.1 \pm 0.4$~m\AA\ at statistical significance of 3.8$\sigma$ (3.1$\sigma$ for the blue and 2.2$\sigma$ for the red) and DR of $1.2 \pm 0.9$.\par

\begin{figure*}[!ht]
      \includegraphics[width=1.\textwidth]{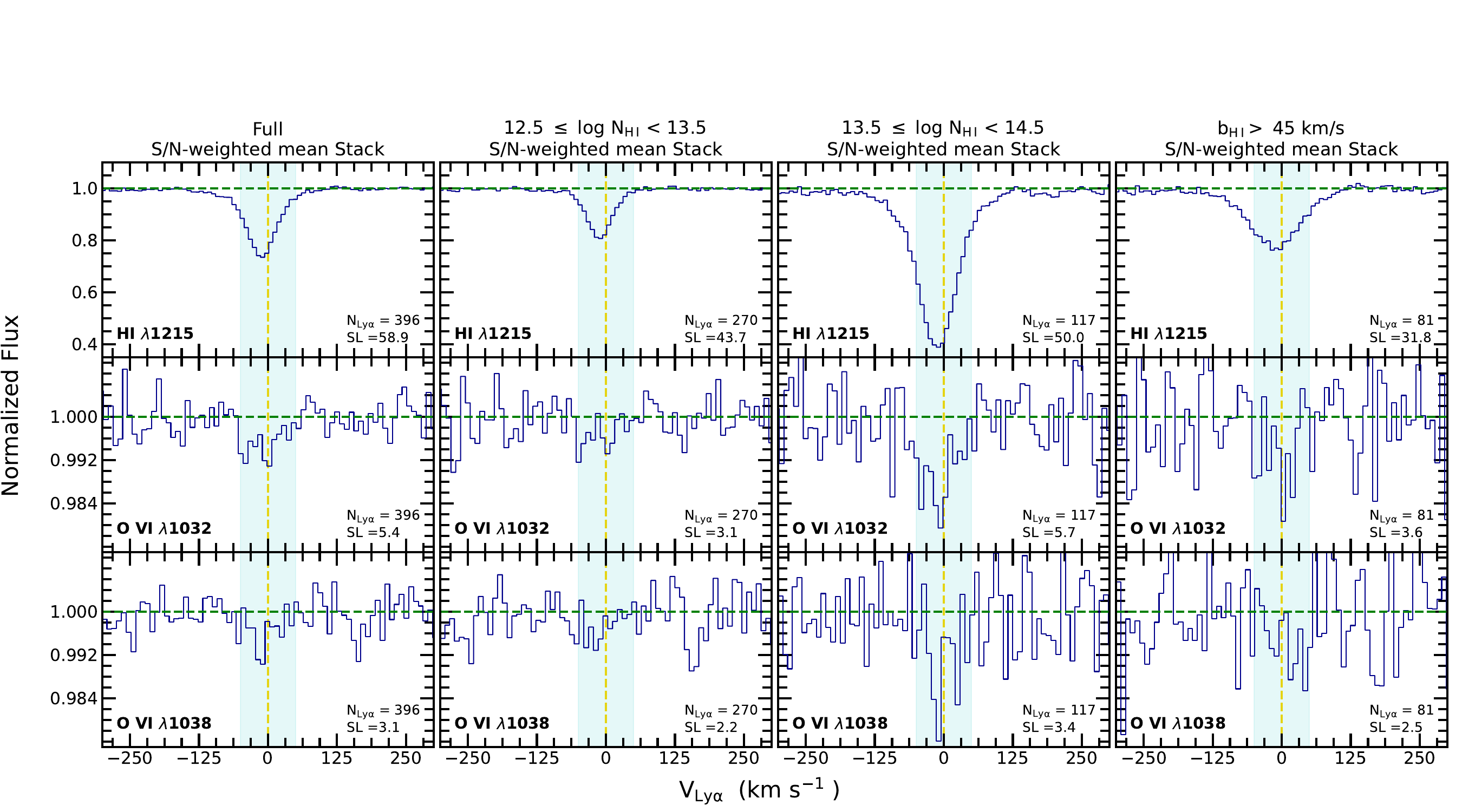}
     \caption{Pseudo-continuum normalized, S/N-weighted mean stacked absorption profiles of \lya~ (top), \ovi~1032 (middle), and \ovi~1038 (bottom) in the velocity space at the rest-frame of the \lya~absorbers. The profiles are shown for four subsamples: the full sample (far-left panels); the low-column density bin (12.5 $\leq$ log$N$ $<$ 13.5; center-left panels); the high-column density bin (13.5 $\leq$ log$N$ $<$ 14.5; center-right panels); and the BLA sample ($b>45$\kms; far-right panels). The shaded regions in each panel encompass the $\pm$50\kms.}
    \label{fig.stack_lya_velocity_space}
\end{figure*}

To compare the kinematics of the detected \ovi\ signal with the stacked \lya\ absorption, we present in Fig.~\ref{fig.stack_lya_velocity_space} pseudo-continuum–normalized, S/N-weighted stacked profiles in velocity space for the full sample (far-left) and the two \hi\ column density bins (center panels). The corresponding stacked \lya\ absorption profiles are shown in the top panels of this figure.\par

As evident from the stacked spectra of all subsamples, the observed \ovi\ doublet lines are well aligned with the corresponding \lya\ absorption. The \lya\ profiles, however, extend over a slightly broader velocity range than the \ovi\ signal, particularly pronounced in the high-column density bin. This becomes more evident when we further subdivide the \lya\ BLA subsample, which are commonly associated with the WHIM, and are characterized by broad \lya\ width with $b > 45$ km s$^{-1}$ \citep{Richter06, Danforth10}. Out of the initial 396 systems, 81 satisfy this BLA criterion. We performed spectral stacking on this subset to search for associated O{\sc vi} absorption. As shown in Fig.~\ref{fig.stack_lya_velocity_space} (far-right panel), we obtain a marginal detection of \ovi\ in this subsample, with $W_{r}^{1032} = 2.2 \pm 0.8$~m\AA, at significance of $4.4\sigma$ (with $3.6\sigma$ and $2.5\sigma$ for the blue and red components, respectively). A larger sample of BLAs is required to confirm the statistical significance of the detected \ovi\ absorption.
The EW of the stacked \lya\ profiles are $78.5 \pm 1.2$ m\AA, $46.8 \pm 1.0$ m\AA, $229.0 \pm 2.9$ m\AA, and $97.0 \pm 2.8$ m\AA\ for the full sample, low-column density, high-column density, and BLA subsample, respectively. While the stacked \ovi\ profiles show two components, the corresponding \lya\ stacks do not show similar structures.\par

\begin{figure}[!ht]
      \includegraphics[width=0.45\textwidth]{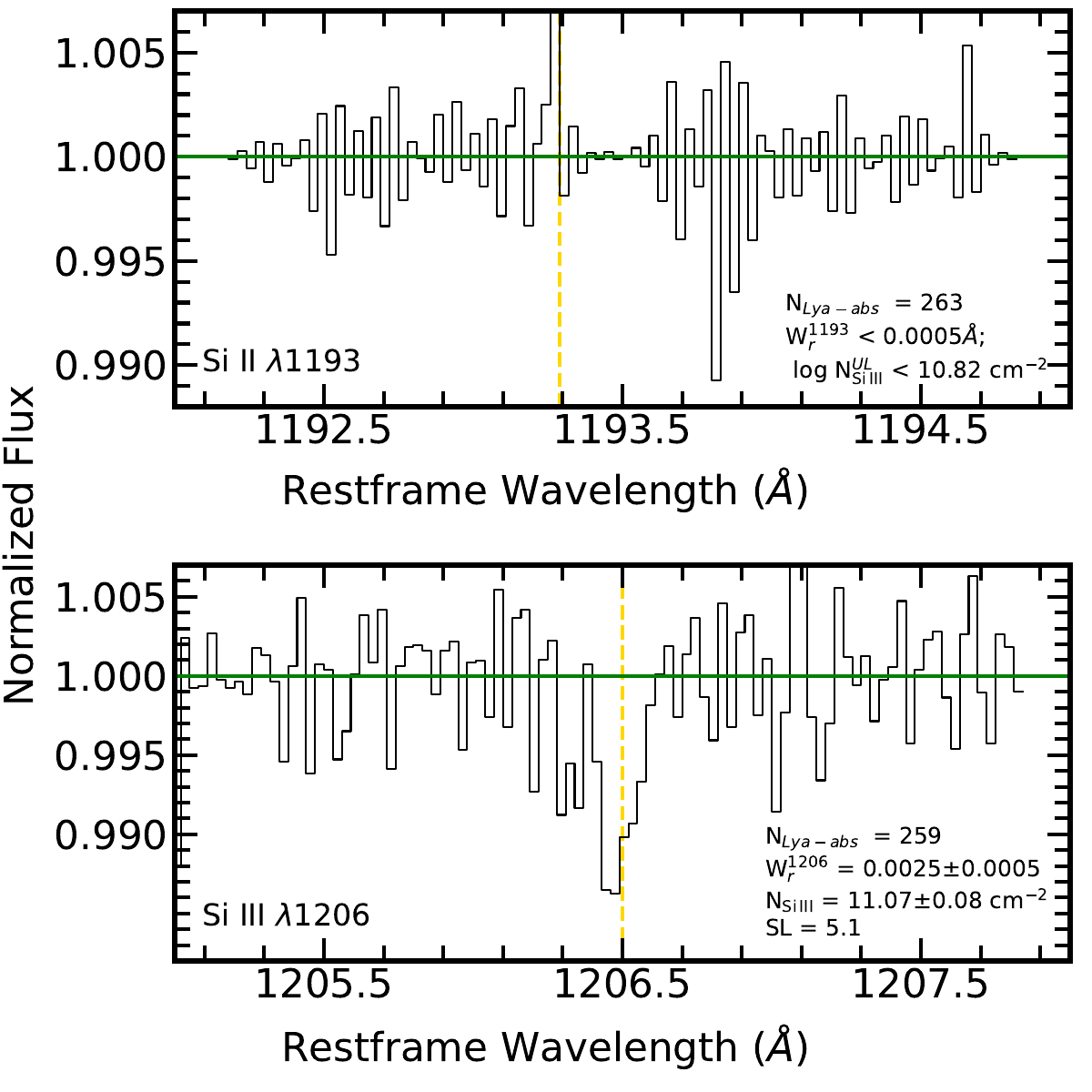}
     \caption{Pseudo-continuum normalized S/N-weighted mean stacks for \siii\ $\lambda$1193 (\emph{top}) and \siiii\ $\lambda$1206 (\emph{bottom}). The number of \lya~ absorbers, the upper limit in the Rest-frame Equivalent Width (REW) and column density is given in the bottom right side of each panel.}
     \label{fig.Si_stack}
    \end{figure}

To explore associated metals alongside our detected \ovi, we focused on the Si~{\sc ii} 
$\lambda 1193$ and Si~{\sc iii} $\lambda 1206$ lines. We selected these ions due to their proximity 
in wavelength to the \lya~lines, thereby ensuring comprehensive coverage of the anticipated spectral 
region within our sample. Following the same contamination removal procedure employed for 
the \ovi~segments, we removed any contamination at the locations corresponding to these ions.
Subsequently, out of the initial 396 systems, we retained 263 systems for Si~{\sc ii} and 259 systems 
for Si~{\sc iii}. As shown in Fig.~\ref{fig.Si_stack}, employing the S/N-weighted mean stacking method for both ions, we obtained 
detection of  Si~{\sc iii} with a significance of $5.1~\sigma$, while no detection for Si~{\sc ii}.
These results provide equivalent width of $2.5 \pm 0.5$~m\AA~for Si~{\sc iii} and an upper limit (3$\sigma$) of $<$ 0.5~m\AA~for Si~{\sc ii}.

We verified that the \nhi\ and $b_{\rm H\,I}$ values obtained by fitting the Voigt profile to the S/N weighted mean \lya~stacks match closely (within 0.05 dex) with the sample mean (Column 4 and 5 of Table~\ref{tab:results}). Therefore, for the subsequent analysis, we adopt the mean \nhi\ and $b_{\rm H\,I}$ values from Table~\ref{tab:results} to estimate the metallicities of \ovi\ absorption in the different subsets.

\begin{table*}
\begin{adjustwidth}{-2cm}{}
\small
\caption{Summary of Measurements Performed on S/N-weighted Mean Stacks} 
\label{tab:results} 
\begin{tabular}{@{}ccccccccccc@{}} 
\hline \hline
Sample  & N$_{\rm Lya}$ & $z_{\rm sys}$  & $<$ log \, $N_{\rm H\,I}$ $>$      & $<b_{\rm H\,I}>$  & W$_{r}^{1032}$   &  SL$_{1032}$  & W$_{r}^{1038}$   & SL$_{1038}$  & log $N_{\rm O\,VI}$   \\
        &                  &   (\AA)        &     (cm$^{-2}$)           &  (\kms)          &  (\AA)           &               &                  &               & (cm$^{-2}$)  \\ 
(1) & (2) & (3) & (4) & (5) & (6) & (7) & (8) & (9)  & (10)  \\
\hline

Full                         & 396 & 0.19 & 13.2 $\pm$ 0.4  & 31.25 & 0.0017 $\pm$ 0.0003 & 5.4 & 0.0013 $\pm$ 0.0003 & 3.1 & 12.14 $\pm$ 0.08 \\
12.5 $\leq$ log$N$ $<$ 13.5  & 270 & 0.19 & 13.1 $\pm$ 0.3 & 29.55 & 0.0011 $\pm$ 0.0004 & 3.1 & 0.0011 $\pm$ 0.0004 & 2.2 & 11.95 $\pm$ 0.15  \\
13.5 $\leq$ log$N$ $<$ 14.5  & 117 & 0.24 & 13.8 $\pm$ 0.3 & 36.9 & 0.0036 $\pm$ 0.0007 & 5.7 & 0.0027 $\pm$ 0.0007 & 3.4 & 12.45 $\pm$ 0.08  \\
BLA ($b>45$\kms)             & 81  & 0.21 & 13.4 $\pm$ 0.4 & 55  & 0.0022 $\pm$ 0.0008 & 3.6 & 0.0021 $\pm$ 0.0008 & 2.5 & 12.25 $\pm$ 0.16 \\
\hline 
\end{tabular} 
\end{adjustwidth}

\begin{tablenotes}\small
\item 
Notes -- (1) Sample name.\\
(2) Number of \lya\ systems.\\
(3), (4), (5) Median redshift, \hi\ column density, and \lya\ $b$-value.\\
(6) REW of \ovi\ 1032 measured within $\pm$50 km s$^{-1}$ from the S/N–weighted mean stacked spectra; the 1$\sigma$ uncertainties are derived from 200 bootstrap realizations.\\
(7) Detection significance of the \ovi\ 1032 line.\\
(8) Same as (6), but for the \ovi\ 1038 line.\\
(9) Same as (7), but for the \ovi\ 1038 line.\\
(10) \ovi\ column density estimated from the 1032 line using the linear part of the curve of growth.
\end{tablenotes}
\end{table*}
%

\section{Discussion}
\label{sec:discussion}

\begin{figure}[!ht]
      \includegraphics[width=0.5\textwidth]{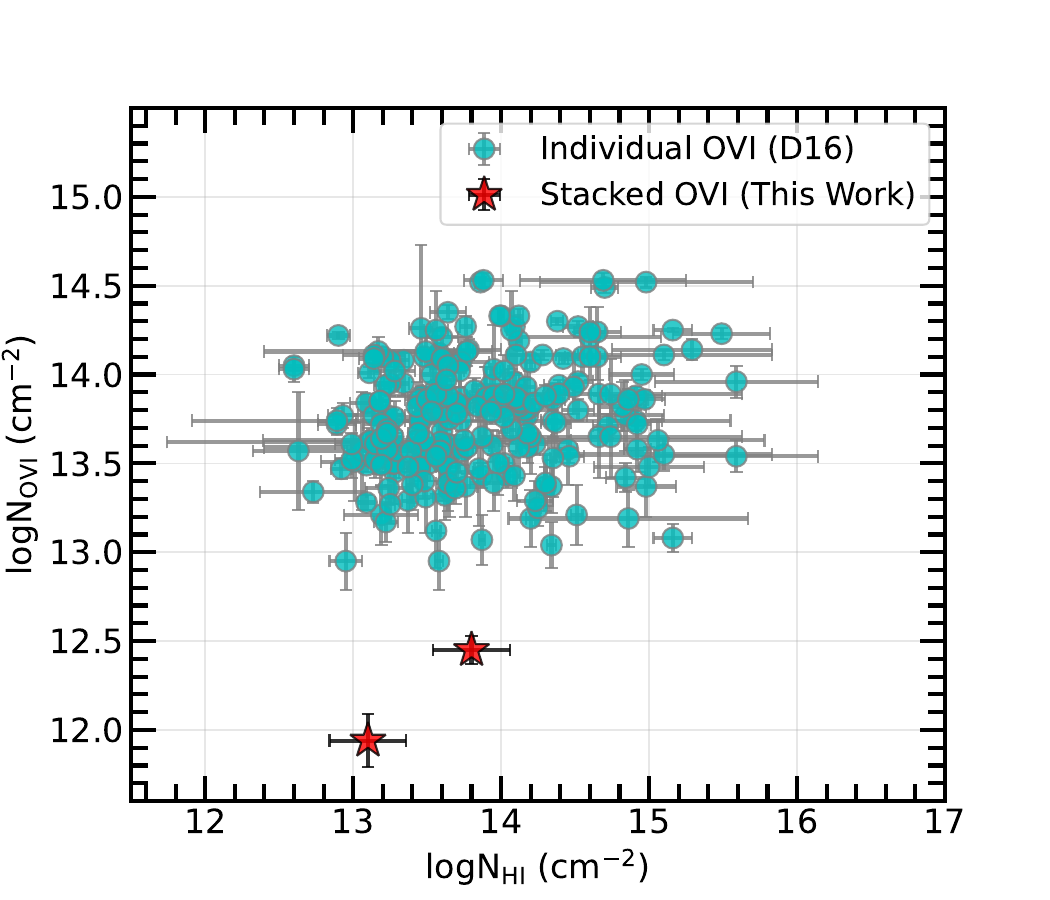}
     \caption{Column densities of \hi\ and \ovi\ for individual absorbers from \citetalias{Danforth16} are shown in cyan circles. The weak \ovi\ absorbers obtained from our stacking analysis for the two bins of log\,$N_{\rm H\,I}$ are shown as red stars.}
     \label{fig:individual_ovi}
    \end{figure}

We detect weak \ovi\ absorption from the low-$z$ IGM at $>5\sigma$ significance by stacking 396 weak ($\log \, N_{\rm HI} < 14.5$) \lya\ absorbers from the \citetalias{Danforth16} sample. 
To compare the column density of our stacked \ovi\ detection with the column densities of individually detected \ovi\ absorbers in the low-$z$ IGM, we again used the \citetalias{Danforth16} catalog, which constitutes the largest low-$z$ IGM metal line survey currently available. We selected all individual \ovi~1032 systems from their sample of 82 AGN sightlines that show \ovi\ detections within $\pm 3\times{\rm FWHM}$ of the $b$-value of the associated \hi\ absorbers, and we retained only those with reliable fitting flags both on \lya\ and \ovi\ lines. This selection yielded 247 individual \ovi\ absorbers. 
Additionally, we excluded six absorbers with column density uncertainties more than 2 dex, resulting in a final sample of 241 individual \ovi\ absorbers from   \citetalias{Danforth16}.

The column densities of these 241 individual \citetalias{Danforth16} \ovi\ absorbers
are shown as a function of the corresponding \hi\ column densities
in Fig.~\ref{fig:individual_ovi} as cyan circles. The column densities derived from our stacked \ovi\ detections for the two log $N_{\rm H\,I}$ bins (see Table~\ref{tab:results}) are shown as red stars at the mean \hi\ column density of each bin. 
Our stacked \ovi\ measurements occupy the lowest region of the $N_{\rm O\,VI}$ parameter space ever probed in the IGM. Whereas CGM \ovi\ measurements tend to cluster around 
log\,$N_{\rm O\,VI} \approx 14.5$ \citep{Heckman2002, fox2011, Tumlinson13, Werk16_ovi}, and the \citetalias{Danforth16} catalog of individual IGM \ovi\ absorbers covers the region $13 < {\rm log}\, N_{\rm O\,VI} < 14.5$,
our stacked \ovi\ signal traces an \emph{entirely new parameter space}, $>$100 times
weaker than the typical CGM \ovi\ signal.
The lowest previously reported \ovi\ column density is $\log \, N_{\rm O\,VI} \approx 12.38$, observed only in the local ISM \citep{Savage2006}. 
In the following subsections, we investigate the galaxy associations and metallicity
of our detected \ovi~absorption.

\subsection{Association with Galaxies}

The \ovi~ion is frequently detected in the CGM of galaxies, 
typically within distances of up to $1-2$ virial radii \citep{Tripp2001,Stocke2006, Wakker2009, Prochaska2011, Tchernyshyov2022, Dutta2025}.
Given the focus of our study on low-column density \lya~absorbers in the search for \ovi, 
we explored whether these systems are situated in proximity to known galaxies. To investigate 
this, we conducted a search in the Sloan Digital Sky Survey (SDSS-IV) archive \citep{Ahumada2020} using 
$CasJobs$\footnote{\color{blue}{https://skyserver.sdss.org/casjobs/}}. 
Of the full sample of 396 \lya\ absorbers toward 57 quasars, 311 absorbers along 44 quasar sightlines fall within the SDSS sky-coverage footprint.

Around these 311 \lya~absorbers, following the approach of \citet{Narayanan2018}, we searched for galaxies with spectroscopic redshifts consistent with each \lya~absorber within a velocity window of $\pm1000$~\kms\ and a projected search radius of 1~Mpc. Within this parameter space, 290 \lya\ absorbers toward 43 quasars show no galaxies within $\pm 1000~\mathrm{km\,s^{-1}}$, while only 21 \lya\ absorbers have associated galaxies, with 30 spectroscopic galaxies with a median magnitude of $r$-mag $= 18.7$. When the projected search radius is increased to 5~Mpc, 212 \lya\ absorbers toward 41 quasars show no associated spectroscopic galaxies. Galaxies were identified around the remaining 99 \lya\ absorbers toward 28 quasars, yielding a total of 279 galaxies in this case.
We do not attempt to include galaxies with photometric redshifts in this analysis. Even the most reliable SDSS photometric redshifts (with ``photoErrorClass'' values of $-1$, $1$, $2$, or $3$) have typical uncertainties of $\Delta z \approx 0.03$ \citep{Beck2016}. At the median redshift of our sample ($z\sim 0.2$), this corresponds to a velocity uncertainty of $\sim7000$~\kms\, equivalent to a projected separation of $\sim100$~Mpc, which is far too large to yield any meaningful constraints on CGM association.
At the median absorber redshift of $z = 0.19$, the SDSS spectroscopic magnitude limit of $r = 17.7$ corresponds to a luminosity threshold of $L \gtrsim 3L^*$. Consequently, our analysis using SDSS is complete only for the brightest galaxies. Within this limitation, 93\% of absorbers have no SDSS spectroscopic galaxy counterparts within 1~Mpc.\par

Further, to extend our search to fainter galaxies, we also used the deep ($g \leq 20$) galaxy survey of \citet{Keeney2018}. The \citet{Keeney2018} survey targets 47 HST/COS sightlines toward UV-bright AGNs and is $\approx$90\% complete down to $g \leq 20$ within 20$^{'}$ of each sightline (see their Table 8), which corresponds to $L \approx 0.15\, L^{*}$ at $z = 0.1$ and $L \approx  L^{*}$ at $z = 0.25$.  
Of the 396 \lya\ absorbers identified along 57 quasar sightlines, 222 \lya\ absorbers towards 28 quasar sightlines are covered by this survey. Around 45 of these 222 \lya\ absorbers, we identify 79 galaxies (with $0.8\, L^{*}$ at a median redshift of 0.13) within 1~Mpc and $\pm 1000$~km~s$^{-1}$ of the \lya\ absorbers. Noting the redshift dependence of the completeness in this survey, we further divide our sample of 222 \lya\ absorbers into two redshift bins: $z \leq 0.25$, where the survey is 90\% complete down to $\approx L^{*}$, and $z > 0.25$. Of these, 127 \lya\ absorbers toward 27 quasars have $z \leq 0.25$, comprising 57\% of our sample, while 95 \lya\ absorbers toward 18 quasars have $z > 0.25$, comprising the remaining 43\%. Among the 127 \lya\ absorbers with $z \leq 0.25$, we find that 43 \lya\ have associated galaxies (76 galaxies in total), while the remaining 84 have no associated galaxies. Of the 95 \lya\ absorbers with $z > 0.25$, only 2 have associated galaxies (3 galaxies in total). It is worth noting, however, that in this redshift bin the survey is complete only for very bright galaxies ($\approx3-4\,L^{*}$ at the bin's median redshift of 0.35). A deeper galaxy survey is therefore needed, particularly for the $z > 0.25$ \lya\ absorbers.\par

This implies that approximately 66\% of \lya\ absorbers in this low-redshift ($z \leq 0.25$) subsample lack an associated galaxy down to $ L^{*}$ within a projected radius of 1~Mpc. This indicates that most absorbers are not associated with luminous galaxies, although the presence of fainter galaxies (with $ < L^{*}$ specially at $z > 0.25$) below the detection limit of the surveys used here cannot be excluded. However, the data presented here suggest that the detected  \ovi~predominantly originates from the diffuse IGM.

\subsection{Metallicity}

In order to measure the metallicity of detected \ovi, we need to understand the ionization mechanism.
If the origin of \ovi~is predominantly photoionization, then the ionization correction depends
on the gas density. However, to obtain density it is usually preferred to have measured
column densities of the ions of the same elements. This is the reason why we searched for Si~{\sc ii}
and Si~{\sc iii}. However, because we do not have clear detection of the Si~{\sc ii}, we resort to 
inferring the gas density from hydrodynamic simulations of the IGM. 

Simulations of the IGM at low redshift have revealed a robust power-law relation between the 
overdensity parameter $\Delta$ and the \lya~column density, 
expressed as $\Delta = \Delta_0 N_{14}^{\gamma}$ \citep[e.g.,][]{Dave10, Smith11, Tepper-Garcia12, Gaikwad17a}, where $N_{14} = N_{\rm HI}/10^{14}~\rm{cm}^{-2}$. 
In this equation, $\Delta$ represents the overdensity, defined as the ratio of the hydrogen number density ($n_H$) to the mean hydrogen density of the IGM 
($\bar{n_H}$) at redshift $z$. The parameters $\Delta_0$ and $\gamma$ denote the normalization and power-law index, respectively. For our analysis, we adopt the values $\Delta_0 = 34.8$ and $\gamma = 0.77$ 
from \citet{Gaikwad17a}, which are consistent with previous 
estimates \citep{Dave10, Smith11}.
These values correspond to $\Delta = 8$ for our entire sample and 
$\Delta = 7$ and $\Delta = 24$ for the binned samples of low and high $N_{\rm H\,I}$, respectively.
Using our adopted cosmology parameters, we determine the mean hydrogen density 
$\bar{n_H}(z) = 1.87 \times 10^{-7}$ cm$^{-3} (1+z)^3$, resulting in log $n_H = -5.6$ 
for the full sample and log $n_H = -5.7$ and $-5.0$ for the binned samples of low and high $N_{\rm H\,I}$, respectively.

With these density values in hand, we perform photoionization equilibrium calculations using the {\sc cloudy} 
code \citep{gunasekera2025}. In our models, we incorporate the extragalactic UV background model developed by \citet{KS19} 
for the ionizing background (their fiducial Q18 model) and apply stopping criteria based on the median \hi~column density of 
the sample, as outlined in Table~\ref{tab:results}. Our analysis yields a metallicity of log\,($Z/Z_{\odot}) = -1.9$  for the full sample 
(i.e, $0.012 Z_{\odot}$). 
Similarly, for the column density binned sample, 
we obtain comparable results, with log\,($Z/Z_{\odot}) = -1.8$ (i.e, $0.016\, Z_{\odot}$) for the low $N_{\rm H\,I}$ bin and $-2.2$ (i.e, $0.006\, Z_{\odot}$) 
for the high $N_{\rm H\,I}$ bin.

Even if we account for the scatter in the $\Delta$--$N_{\rm HI}$ relation, 
expressed as $\Delta = (\Delta_0 \pm d\Delta_0)\, N_{14}^{\gamma \pm d\gamma}$, 
the resulting relative error in the overdensity is
\begin{equation*}
    \frac{d\Delta}{\Delta} = \sqrt{\left(\frac{d\Delta_0}{\Delta_0}\right)^2 
    + \left(d\gamma \, \ln N_{14}\right)^2}.
\end{equation*}
Using $\Delta_0 = 34.8$, 
$d\Delta_0 = 5.9$, and $d\gamma = 0.22$ from \citet{Gaikwad17a}, 
this yields relative errors of $\sim 44\%$, $\sim 49\%$, and $\sim 20\%$ 
for the full sample, the low column density bin  $ 12.5 < \log N_{\rm HI} < 13.5$ 
and high column density bin $ 13.5 < \log N_{\rm HI} < 14.5$, respectively (see Table~\ref{tab:results} for $N_{14}$ values). 
These overdensity uncertainties propagate directly into our derived $n_{\rm H}$ whereas the $n_{\rm H}$  uncertainties propagate into the derived metallicities through the
ionization fractions, since $Z \propto (N_{\rm OVI}/N_{\rm HI})\,(f_{\rm HI}/f_{\rm OVI})$.
For the two lower-density cases ($\log n_{\rm H}=-5.6$ for the full sample and
$\log n_{\rm H}=-5.7$ for the low-column-density bin), which lie below the
O~{\sc vi} photoionization peak, 
both $f_{\rm HI}$ and $f_{\rm OVI}$ increase with
$n_{\rm H}$, so their density dependence largely cancels in the ratio; the
$\sim$44--49\% uncertainties in $n_{\rm H}$ then translate into only
$\sim$5--10\% uncertainty in $Z$. For the high-column-density bin,
$\log n_{\rm H}=-5$ lies close to the broad peak in $f_{\rm OVI}$, where
$f_{\rm OVI}$ is nearly flat and this cancellation weakens; the 20\% uncertainty
in $n_{\rm H}$ then translates into a 24\% uncertainty in $Z$. 
In all cases, the
induced uncertainty is $\lesssim$0.1~dex, and our metallicity values 
should be interpreted with these uncertainties in mind.

On the other hand, if the \ovi~originates in gas in 
collisional ionization equilibrium (CIE) at
$T =  10^{5.5}$\,K, where the fraction of \ovi~gas 
is at its highest \citep{Sutherland1993, Gnat2007G},
the metallicity of gas is given by the expression 
${\rm log}\, (Z/Z_{\odot}) = {\rm log}\, N_{\rm O\,VI} - {\rm log}\, N_{\rm H\,I} -5.22 +3.31$, where the number 
$-3.31$ is the abundance (O/H) of Sun \citep{Gass10}, 
and $5.22$ is the log of the ratio of ionization 
fraction of \ovi~and \hi\ in a CIE model at $10^{5.5}$\,K. 
Using this expression, we find that,
if \ovi~traces same \hi~column density 
as the median of \lya~sample, then the metallicity of \ovi~bearing gas is $\approx0.001\,Z_{\odot}$. 
This inferred metallicity is highly temperature-sensitive and can vary from $\simeq 2$ to $10^{-4}\,Z_{\odot}$ 
for $T=10^{5-6}$ K. We adopt $T=10^{5.5}$ K because this temperature corresponds 
to the peak \ovi\ ion fraction in CIE of 0.22 \citep{Gnat2007G}. 
The resulting metallicity should therefore be regarded as 
a lower limit, since not all of the observed \ovi\ necessarily arises from collisionally 
ionized gas. Additional uncertainties can arise because the measured column of H~{\sc i}
may not trace gas with the 
same temperature, kinematics, or line width as the \ovi-bearing phase, and the adopted gas temperature 
may not be consistent with CIE conditions. \par

The metallicity associated with the lowest column density \lya\ absorbers ($\log \, N_{\rm H\,I} < 14$) at $z < 0.5$ is key probe of the chemical enrichment of the diffuse IGM, but remains poorly constrained due to observational limitations. These absorbers arise in low-density regions of the cosmic web ($\Delta \approx 10$), where metal lines are intrinsically faint and fall below the detection thresholds of individual UV spectra.
\citetalias[][]{Danforth16} survey showed that the fraction of metal-bearing absorbers drops steeply with decreasing $N_{\rm HI}$: for ${\rm log} \, N_{\rm HI}< 13.5$, only $\sim 3$\% of systems show detectable metals, rising to $\sim 22$\% for $13.5 < {\rm log}\, N_{\rm HI} < 14.5$, and becoming nearly ubiquitous only for ${\rm log}\ N_{\rm HI} > 14.5$. This trend is attributed both to instrumental sensitivity limits and to genuinely lower metallicities in the most diffuse IGM. 
Cosmological simulations predict that the under-dense $z \approx 0$ IGM should contain only trace metals in the range of $\approx 10^{-3} - 10^{-2}\,Z_\odot$, with the scatter in the estimate due to differences in the feedback model and local enrichment history \citep{Aguirre2001, Shen2010, Cen2011}.

As pointed out earlier, low-redshift \ovi~surveys have been sensitive only to $\log \, N_{\rm O\,VI} \geq 13$ \citep{Tripp08, Danforth2008, Savage14}, corresponding to much stronger \lya~systems potentially associated with the CGM gas. Stacking has probed the average metal absorption, associated with lower  $N_{\rm HI}$ systems, that are individually undetectable. Our measurements provide, for the first time, observational evidence of metals even in the weakest \lya~ absorbers at  $z < 0.5$, placing important constraints on the metal-enriched fraction of the diffuse IGM.

The spectral stacking method demonstrated here offers a pathway to probing any potential metallicity floor in the low-redshift IGM. Identifying such a floor would provide valuable insight into when and how the IGM was enriched by galactic outflows, and would place important constraints on feedback processes in galaxy formation \citep[see e.g.,][]{Simcoe2004, Schaye2007}. 
Our photoionization modeling resulted in a lowest metallicity estimate of $0.006 Z_{\odot}$ when we stacked \ovi~in our subsample of H~{\sc i} with $13.5< {\rm log}~{N_{\rm H\,I}} <14.5$ whereas we find $0.016 Z_{\odot}$ for subsample of H~{\sc i} with $12.5< {\rm log}~{N_{\rm H\,I}} <13.5$.  Simulations also show that lower column density systems can have metallicities well below $0.01 Z_{\odot}$, consistent with accretion of metal poor IGM gas onto galaxies \citep{Hafen2017, Rahmati2018}.
There is no evidence for a universal metallicity floor at or above $0.01 Z_{\odot}$ but this value is consistent with the mean metallicity inferred for the low-$z$ IGM \citep[see][]{Prochaska17_metals, Oppenheimer2012, Shull2014}. Taken together, our results highlight that while the mean metallicity of the low-redshift IGM is of the order $0.01 Z_{\odot}$ obtained with the assumptions of photoionization origin of detected O~{\sc vi} gas, the absence of a detectable lower bound in our stacked measurements underscores that the true metallicity floor, if it exists, must lie below this level.

\section{Conclusions}\label{sec.results}

In our study, we employed a spectral stacking method to unveil the presence of 
\ovi~absorption lines within the low-column density \lya~absorbers. Our dataset consisted 
of 82 high S/N quasar spectra from HST-COS, probing the IGM at $z<0.5$. 
Our initial screening aimed to identify \lya~absorbers that lacked any individual 
detections of \ovi\ (so have log\,$N_{\rm O\,VI}\lesssim13$). We successfully identified 396 such absorbers at
$0.1 < z < 0.5$ and confirmed that they were uncontaminated at the expected \ovi~locations.

Using this set of 396 absorbers, we carried out spectral stacking through three distinct methods, 
leading to the detection of \ovi~doublets (Fig.~\ref{fig.sample_stack}). These detections show equivalent width W$^{1032}_{r}$ of $1.7\pm0.3$m\AA~ at significance level of 6.2$\sigma$ with a corresponding \ovi~column density of log $N_{\rm O\,VI} = 12.14 \pm 0.08$ in the S/N-weighted mean stack spectrum, around $\approx$1\,dex weaker than any individual detectable \ovi\ component.

Upon dividing our absorber sample into two bins based on \hi~column density, 
low-column density ( $12.5 \leq$ log$N_{\rm H\,I} < 13.5$ ) and high-column density ($13.5 \leq$ log\,$N_{\rm H\,I}< 14.5$ ), 
we found that the \ovi\ absorption EW in the high-column-density bin is stronger than in the low-column-density bin by more than $3\sigma$.

Out of these 396 systems, we created a sample of potential broad \lya~alpha absorbers 
containing 81 \lya~lines having $b_{\rm H\,I}> 45$\kms\ and performed spectral stack 
to detect \ovi. Our search yields a marginal detection (Fig.~\ref{fig.stack_lya_velocity_space}, far-right panels) of 
\ovi\ with W$_{r}^{1032}$ of $2.2 \pm 0.8$~m\AA\ at 4.4$\sigma$ and and column density of 
log\,$N_{\rm O\,VI} = 12.25 \pm 0.16$. A larger BLA sample is needed to confirm this result. Our results are summarized in Table~\ref{tab:results}. 

We further investigated the presence of other metal absorbers associated to 
\ovi~focusing on stacking \siii\ and \siiii\ lines. Our findings included no
detection of Si~{\sc ii},  and a 5$\sigma$ detection for Si~{\sc iii}, 
revealing an equivalent width of 2.5 $\pm$ 0.5 m\AA~and a 
corresponding column density of  log\,$N_{\rm Si\,III} = 11.07 \pm 0.08$ (Fig.~\ref{fig.Si_stack}).

We confirm that 93\% of our \lya~absorbers do not trace any bright galaxies ($L \lesssim 3L^*$) within 1~Mpc distance.
A more complete search of the low-redshift subsample shows that 66\% lack an associated galaxy down to $L^\ast$ for $z \leq 0.25$.
Hence the \ovi~absorption reported here is likely originating from the low density IGM. Future deep surveys will be essential to further confirm whether the detected \ovi~is not associated with the CGM of faint galaxies (particularly $ \lesssim L^{*}$ at $z > 0.25$) below the completeness of the surveys.
The overdensities inferred for these weak \ovi~ systems ($\Delta \approx 7-24$ ) place them in the diffuse to mildly overdense regions of large-scale filaments.

In terms of gas metallicity,
the detected \ovi\ traces gas with an average metallicity of $\approx 0.01 \, Z_\odot$ in the case of photoionization, and $\approx 0.001\, Z_{\odot}$ in the case of 
collisional ionization at $10^{5.5}$\,K in the WHIM. This study probes the average metal absorption associated with low–column density \lya\ systems that are individually undetectable, providing the first observational evidence of metals in the weakest \lya~ absorbers at $z < 0.5$ and placing important constraints on the metal-enrichment of the underdense IGM and a possible metallicity floor.

\vspace{0.2 cm}
{\it Acknowledgments:}

We thank the referee for the constructive feedback.
This research is based on observations made with the NASA/ESA Hubble Space Telescope obtained from the Space Telescope Science Institute, which is operated by the Association of Universities for Research in Astronomy, Inc., under NASA contract NAS 5–26555. VK thanks the Inter-University Centre for Astronomy and Astrophysics (IUCAA), Pune, India, for providing research facilities and support under the IUCAA Associateship Programme.


All of the HST data presented in this article were obtained from the Mikulski Archive for Space Telescopes (MAST) at the Space Telescope Science Institute. The specific observations analyzed can be accessed via \dataset[DOI: 10.17909/ccvq-me89]{https://doi:10.17909/ccvq-me89}.

\appendix

\setcounter{section}{0}
\setcounter{table}{0}
\setcounter{figure}{0}
\renewcommand{\thefigure}{A\arabic{figure}}
\renewcommand{\thetable}{A\arabic{table}}  
\renewcommand{\thesection}{A\arabic{section}}

\section{Stacking Analysis After Removing Uncorrelated Absorption}

\begin{figure*}[!ht]
      \includegraphics[width=1.0\textwidth]{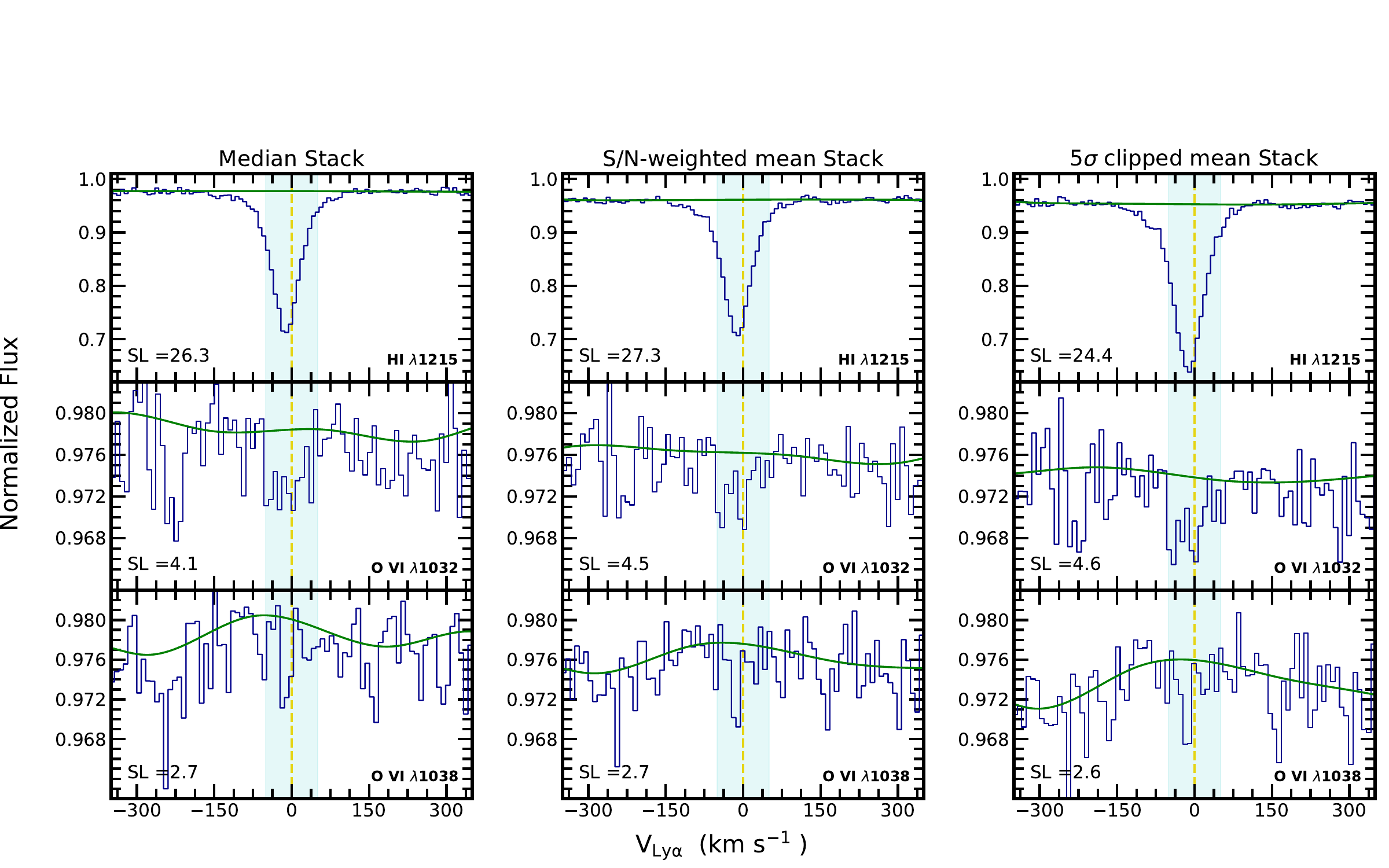}
     \caption{Stacked absorption profiles of \lya~ (top), \ovi-1032 (middle), and \ovi-1038 (bottom) in the velocity space at the rest-frame of the \lya~absorbers. The profiles are obtained through three methods: median (left panels), S/N-weighted mean (middle panels), and 5$\sigma$ clipped mean (right panels). The shaded regions encompass the $\pm$50\kms~ around absorption features. These stacked profiles are generated by applying 3$\sigma$ clipping to uncorrelated random absorbers within a velocity interval of $\pm$[100, 1000]\kms~ around each absorption feature, resulting in the suppression of the emission bump observed in the original stacked profiles (see Fig.~\ref{fig.sample_stack}). }
    \label{fig.appendix_sample_stack_no_bump}
\end{figure*}

In this section, we demonstrate that the emission-like feature observed in the original stacked spectra has no physical origin and instead arises from our sample selection, which preferentially includes spectra with no absorption within $\pm$50 km s$^{-1}$ of the \ovi\ doublet. To assess the impact of excluding random, uncorrelated absorption outside our selection windows, we extracted spectral regions spanning $-$1000 to $-$100 \kms~ on the blue side and 100 to 1000 \kms~ on the red side around both expected locations of the \ovi~ doublet. We use a $\pm$100 km s$^{-1}$ window, rather than $\pm$50 km s$^{-1}$, to ensure that we do not manually exclude any \ovi~ absorbers that may be offset from their associated \lya\ absorbers by larger velocities. We then applied 3$\sigma$ clipping to remove significant random absorption features and replaced the clipped pixels by interpolating using the 1$\sigma$ rms continuum fluctuations of each quasar spectrum. The stacking analysis was repeated using the median, S/N-weighted mean, and 5$\sigma$-clipped mean for the full sample as well as for various subsamples. The resulting stacks for the full sample, shown in velocity space for both \lya\ and the \ovi~ doublet (Fig.~\ref{fig.appendix_sample_stack_no_bump}), exhibit no emission-like features. Instead, we detect significant \ovi\ absorption in all stacking methods: 5.0$\sigma$ (4.1$\sigma$ for \ovi\ 1032 and 2.7$\sigma$ for \ovi\ 1038) for the median stack; 5.2$\sigma$ (4.5$\sigma$ for \ovi\ 1032 and 2.7$\sigma$ for \ovi\ 1038) for the S/N-weighted mean stack; and 5.3$\sigma$ (4.6$\sigma$ for \ovi\ 1032 and 2.6$\sigma$ for \ovi\ 1038) for the 5$\sigma$-clipped mean stack. The measured parameters for these revised stacks, for both the full sample and the subsamples, are listed in Table~\ref{tab:appendix_results} for the S/N-weighted mean stacks.

\label{appendix:section_remove_uncor}

\begin{table*}
\begin{adjustwidth}{-2cm}{}
\small
\caption{Summary of Measurements Performed on S/N-weighted Mean Stacks Excluding the Uncorrelated Absorption} 
\label{tab:appendix_results} 
\begin{tabular}{@{}ccccccccccc@{}} 
\hline \hline
Sample  & N$_{\rm Lya}$ & $z_{\rm sys}$  & $<$ log \, $N_{\rm H\,I}$ $>$      & $<b_{\rm H\,I}>$  & W$_{r}^{1032}$   &  SL$_{1032}$  & W$_{r}^{1038}$   & SL$_{1038}$  & log $N_{\rm O\,VI}$   \\
        &                  &   (\AA)        &     (cm$^{-2}$)           &  (\kms)          &  (\AA)           &               &                  &               & (cm$^{-2}$)  \\ 
(1) & (2) & (3) & (4) & (5) & (6) & (7) & (8) & (9)  & (10)  \\
\hline

Full                        & 396 & 0.19 & 13.2 $\pm$ 0.44 & 31.25 & 0.0013 $\pm$ 0.0004 & 4.5 & 0.0009 $\pm$ 0.0003 & 2.7 & 12.02 $\pm$ 0.12  \\  
12.5 $\leq$ log$N$ $<$ 13.5 & 270 & 0.19 & 13.1 $\pm$ 0.26 & 29.55 & 0.0012 $\pm$ 0.0004 & 2.6 & 0.0011 $\pm$ 0.0004 & 2.7 & 11.97 $\pm$ 0.15  \\ 
13.5 $\leq$ log$N$ $<$ 14.5 & 117 & 0.24 & 13.8 $\pm$ 0.25 & 36.9  & 0.0037 $\pm$ 0.0007 & 7.9 & 0.0027 $\pm$ 0.0007 & 3.6 & 12.47 $\pm$ 0.08   \\  
BLA ($b>45$\kms)            & 81 & 0.21  & 13.4 $\pm$ 0.36 & 55.4  & 0.0022 $\pm$ 0.0008 & 2.9 & 0.0014 $\pm$ 0.0008 & 2.0 & 12.24 $\pm$ 0.16    \\

\hline 
\end{tabular} 
\end{adjustwidth}

\begin{tablenotes}\small
\item 
Notes -- (1) Sample name.\\
(2) Number of \lya\ systems.\\
(3), (4), (5) Median redshift, \hi\ column density, and \lya\ $b$-value.\\
(6) REW of \ovi\ 1032 measured within $\pm$50 km s$^{-1}$ from the S/N–weighted mean stacked spectra; the 1$\sigma$ uncertainties are derived from 200 bootstrap realizations.\\
(7) Detection significance of the \ovi\ 1032 line.\\
(8) Same as (6), but for the \ovi\ 1038 line.\\
(9) Same as (7), but for the \ovi\ 1038 line.\\
(10) \ovi\ column density estimated from the 1032 line using the linear part of the curve of growth.
\end{tablenotes}
\end{table*}

\bibliography{Reference}{}
\bibliographystyle{aasjournal}

\end{document}